\documentclass[sigconf]{acmart}
\usepackage{graphicx}
\usepackage{algorithm}
\usepackage{rotating}
\usepackage{balance}

\usepackage{tabularx}
\usepackage{subfigure}
\usepackage{bm}
\usepackage{multirow}
\usepackage{enumitem}

\AtBeginDocument{%
  \providecommand\BibTeX{{%
    \normalfont B\kern-0.5em{\scshape i\kern-0.25em b}\kern-0.8em\TeX}}}

\setcopyright{acmlicensed}
\copyrightyear{2018}
\acmYear{2018}
\acmDOI{XXXXXXX.XXXXXXX}

\acmConference[DASFAA '25]{The 30th International Conference on Database Systems for Advanced Applications}{May 26-29, 2025}{Singapore}
\acmBooktitle{The 30th International Conference on Database Systems for Advanced Applications (DASFAA ’25)}
\acmISBN{978-1-4503-XXXX-X/18/06}

\begin{document}

\title[AdaF$^2$M$^2$: Comprehensive Learning and Responsive Leveraging Features in Recommendation System]{AdaF$^2$M$^2$: Comprehensive Learning and Responsive Leveraging Features in Recommendation System}

\author{Yongchun Zhu}
\affiliation{%
  \institution{ByteDance}
  \city{Beijing}
  \country{China}}
\email{zhuyc0204@gmail.com}

\author{Jingwu Chen$^{\dag}$}
\affiliation{%
  \institution{ByteDance}
  \city{Beijing}
  \country{China}}
\email{chenjingwu@bytedance.com}

\author{Ling Chen}
\affiliation{%
  \institution{ByteDance}
  \city{Shanghai}
  \country{China}}
\email{chenling.nb@bytedance.com}

\author{Yitan Li}
\affiliation{%
  \institution{ByteDance}
  \city{Shanghai}
  \country{China}}
\email{liyitan@bytedance.com}

\author{Feng Zhang}
\affiliation{%
  \institution{ByteDance}
  \city{Shanghai}
  \country{China}}
\email{feng.zhang@bytedance.com}

\author{Xiao Yang}
\affiliation{%
  \institution{ByteDance}
  \city{Beijing}
  \country{China}}
\email{wuqi.shaw@bytedance.com}

\author{Zuotao Liu}
\affiliation{%
  \institution{ByteDance}
  \city{Shanghai}
  \country{China}}
\email{michael.liu@bytedance.com}

\thanks{
$\dag$ Jingwu Chen is the corresponding author.}

\renewcommand{\shortauthors}{Zhu, et al.}

\begin{abstract}
Feature modeling, which involves feature representation learning and leveraging, plays an essential role in industrial recommendation systems. However, the data distribution in real-world applications usually follows a highly skewed long-tail pattern due to the popularity bias, which easily leads to over-reliance on ID-based features, such as user/item IDs and ID sequences of interactions. Such over-reliance makes it hard for models to learn features comprehensively, especially for those non-ID meta features, e.g., user/item characteristics. Further, it limits the feature leveraging ability in models, getting less generalized and more susceptible to data noise. Previous studies on feature modeling focus on feature extraction and interaction, hardly noticing the problems brought about by the long-tail data distribution. To achieve better feature representation learning and leveraging on real-world data, we propose a model-agnostic framework AdaF$^2$M$^2$, short for \textbf{A}daptive \textbf{F}eature \textbf{M}odeling with \textbf{F}eature \textbf{M}ask. The feature-mask mechanism helps comprehensive feature learning via multi-forward training with augmented samples, while the adapter applies adaptive weights on features responsive to different user/item states. By arming base models with AdaF$^2$M$^2$, we conduct online A/B tests on multiple recommendation scenarios, obtaining +1.37\% and +1.89\% cumulative improvements on user active days and app duration respectively. Besides, the extended offline experiments on different models show improvements as well. AdaF$^2$M$^2$ has been widely deployed on both retrieval and ranking tasks in multiple applications of Douyin Group, indicating its superior effectiveness and universality. 

\end{abstract}




\maketitle

\section{Introduction}

Video platforms like Douyin and TikTok are extremely hot nowadays, with billions of users. 
The progress of personalized recommendation systems has made a significant contribution to the success of those platforms, which helps expose attractive items to users according to their interests.  
While tracing the source of personalization, features play an essential role by providing the original information of users, items, and interactions between them. 
Actually, feature engineering has been very important in industrial recommendation systems from past to present. 

Conventional matrix factorization (MF) methods~\cite{singh2008relational,koren2009matrix,rendle2009bpr} work by decomposing the user-item interaction matrix into the product of two lower dimensional matrices, where each row represents a user/item ID embedding. 
After MF, Factorization Machines (FMs)~\cite{rendle2010factorization} leverage more features in addition to ID and conduct pair-wise interactions via the inner product of two embeddings. 
With more computing firepower and the growing scale of data, recommendation models in industry have migrated from FM-based models to Deep Neural Networks (DNN)~\cite{covington2016deep,zhang2016collaborative,zhou2018deep,bhagat2018buy,lian2018xdeepfm,he2017neural,zhang2024uncovering,zhu2024interest}, which significantly improve model performance. 
DNNs have the ability to model more complex feature interactions~\cite{cheng2020adaptive,wang2021dcn,tian2023eulernet}, thus being able to handle more types of feature inputs, e.g., image~\cite{wei2019mmgcn,sun2020multi} and text~\cite{chen2019follow}. 
Reviewing the development history of recommendation models, the utilization of features is getting better, and more complex features are becoming acceptable, varying from ID to multimodal features. 

In practice, recommendation models usually take lots of features as inputs, such as IDs, sequences, and characteristics. 
With diverse features engaged in, feature modeling is crucial for final performance. 
Well-learned feature representations can help models generalize better, handle noise and uncertainty in the data, and uncover hidden patterns. 
However, the data distribution in industrial recommendation systems usually follows a highly skewed long-tail pattern due to the popularity bias, e.g., 5\% of users/items account for over 80\% of samples. 
Head users probably dominate the model learning with a large proportion of samples. 
Since the user/item ID is the most fine-grained and the most informative feature for users with a large amount of behavioral data, the model also tends to attribute more knowledge to IDs. 
This easily leads to over-reliance on ID-based features, such as user/item IDs and ID sequences of interactions. 
Such over-reliance will seriously affect the comprehensive learning of features, making the model less generalized and more vulnerable to noise interference. 
Taking data noise as an example, user behavior is often interfered with unobserved factors. For instance, a user might skip a favorite video just because they received a phone call at that time, but such noisy pattern can be wrongly fitted by the user ID. 

Further, the above over-reliance also limits the learning of non-ID meta features, causing poor performance in the user/item cold-start stage. 
For example, recommending a new video with few samples highly relies on its attribute features such as theme and category. 
 The weakening of feature representation learning will make the feature leveraging less responsive to different user/item states. 

In this paper, to tackle problems brought about by the long-tail distribution, we propose a simple but powerful framework Adaptive Feature Modeling with Feature Mask (\textbf{AdaF$^2$M$^2$}), which consists of a feature-mask mechanism and a state-aware adapter. We generate multiple augmented samples with the feature-mask mechanism and learn features comprehensively via multi-forward task-oriented training on these samples. To adaptively model users/items of different states, we propose the state-aware adapter, which takes empirical state signals as input, to apply adaptive weights on features responsive to different user/item states. Note that it is convenient to deploy AdaF$^2$M$^2$ with different base recommendation models.

The main contributions of our work are summarized into four folds:
\begin{itemize}
    \item To achieve feature representation comprehensive learning and responsive leveraging on real-world data, we propose a simple but powerful framework AdaF$^2$M$^2$. 
    \item We conduct multiple rounds of online experiments on multiple recommendation scenarios, obtaining +1.37\% and +1.89\% cumulative improvements on user active days and app duration respectively.
    \item We conduct offline experiments on both public datasets and industrial datasets with billions of samples to demonstrate the effectiveness of AdaF$^2$M$^2$. In addition, the offline results testify that it is convenient to apply the proposed framework upon different base models.
    \item AdaF$^2$M$^2$ has been widely deployed on both retrieval and ranking tasks in multiple applications of Douyin Group, indicating its superior effectiveness and universality.
\end{itemize}

\section{Related Work}
In this section, we will introduce the related work of recommendation systems from two aspects: Feature Representation Learning and Feature Representation Leveraging.

\textbf{Feature Representation Learning.} Researchers have investigated feature representation learning for a long time. 
Feature interaction aims at capturing associations among different features and producing more diverse combinations of features to improve feature representations. The early works focus on how to produce cross-feature interaction. Factorization machines (FM)~\cite{rendle2010factorization} have been shown to be an effective method for feature interaction, and many methods~\cite{juan2016field,guo2017deepfm,he2017neural,lian2018xdeepfm,sun2021fm2} capture high-order interaction based on FM. In addition, some methods~\cite{wang2017deep,wang2021dcn} exploit Multi-layer Perceptron to produce high-order interactions. 
However, these methods generally enumerate all the cross features under a predefined maximum order, which is very time-consuming, so some researchers pay attention to adaptively feature interactions. With the advantage of attention mechanisms, multi-head attention is utilized for adaptively feature interactions~\cite{huang2019fibinet,song2019autoint,zhang2021multi}. In addition,  
AutoML is exploited to seek useful high-order feature interactions without manual feature selection~\cite{liu2020autogroup}. 
Due to the high computational costs caused by the exponential growth of high-order feature combinations, some researchers pay attention to efficient feature interaction~\cite{deng2021deeplight,tian2023eulernet,zhang2023lightfr}, which aims at reducing the computational costs. 
Besides, many methods focus on improving representations of some specific features, e.g., item ID~\cite{pan2019warm,ouyang2021learning,zhu2021learning}, user ID~\cite{man2017cross,zhu2022personalized,cao2022disencdr,cao2022cross}, short-term sequential features~\cite{tang2018personalized,zhou2018deep,zhang2021talent,wu2022multi,xin2025llmcdsr,zhu2024interest,sun2024large}, long-term sequential features~\cite{pi2020search,chang2023twin}, multi-modal features~\cite{wei2019mmgcn,sun2020multi,zhang2023interactive}. 

In addition, some researchers make efforts to enhance representations with novel training frameworks. DropoutNet~\cite{volkovs2017dropoutnet} applies dropout to ID embeddings, which improves the recommendation performance for cold-start users and items. Yao et al.~\cite{yao2021self} exploits self-supervised learning to improve item representation learning as well as serve as additional regularization to improve generalization. 
However, these methods ignore the importance of effective feature representation leveraging for users/items of different states and lack additional supervised signals.

\textbf{Feature Representation Leveraging.} 
The useless features may introduce noise and complicate the training process~\cite{sun2021discerning,sun2021market}. Feature representation leveraging aims at identifying useful feature interactions through model training. The attention mechanism is convenient to model importance of features, and many methods~\cite{xiao2017attentional,huang2019fibinet,song2019autoint,zhang2021multi} utilize it for the feature selection. AFN~\cite{cheng2020adaptive} takes learnable parameters for feature selection. AutoFIS~\cite{liu2020autofis} exploited a two-stage framework, which identifies the feature importance and retrains the model without redundant feature interactions. However, these methods detect beneficial features by learning from the features themselves, which could be dominated by the head users and items, and the ID-based features are usually assigned with high weights. In this paper, we propose a state-aware adapter taking empirical state signals as input responsive to different user/item states.

\vspace{-0.3cm}
\section{Proposed Framework}

In this section, we introduce our proposed framework named Adaptive Feature Modeling with Feature Mask (AdaF$^2$M$^2$), which 
consists of a feature-mask mechanism and a state-aware adapter. In Section~\ref{sec:3.1}, we specify the common setup of a recommendation task in industrial recommendation systems. In Section~\ref{sec:fm}, we explain the feature-mask mechanism and how it helps comprehensive feature learning. In Section~\ref{sec:3.3}, we present a state-aware adapter that is able to apply adaptive weights on features responsive to different user/item states. In Section~\ref{sec:3.4}, we demonstrate the overall framework and how to deploy it in ranking and retrieval models. 


\subsection{Recommendation Task Setup}\label{sec:3.1}
First, we consider the common setup for a binary classification task, such as CTR predicting in recommendation systems. Each sample consists of the input raw features $\bm{x} = [x_1, \cdots, x_n]$ and a label $y \in \{0,1\}$, where $n$ indicates the number of raw features. 
In binary classification, a deep recommendation model approximates the probability $\hat{y} = Pr(y=1|\bm{x})$ for the sample with input $\bm{x}$. Generally, a deep recommendation model consists of a feature embedding layer, a feature interaction layer, and a deep network. The feature embedding layer aims to transform raw features $[x_1, \cdots, x_n]$ into low-dimensional representations, named feature embeddings, denoted as $[\bm{v}_1, \cdots, \bm{v}_n]$. Then, the feature interaction layer takes the embeddings as input to generate high-order cross-feature representations, and most existing methods focus on this layer, e.g., FM~\cite{rendle2010factorization}, DCN~\cite{wang2021dcn}. Finally, the cross-feature representations are fed into a deep network for the prediction. In this paper, we focus on improving the representation of the embedding layer, and the interaction layer and the deep network are denoted as $g(\cdot)$. The prediction of a recommendation model is formulated as:
\begin{equation}
    \hat{y} = g([\bm{v}_1, \cdots, \bm{v}_n]).
\end{equation}

The cross-entropy loss is often used as the optimization target for binary classification:
\begin{equation}
    \mathcal{L} = -y \log \hat{y} - (1-y) \log (1 - \hat{y}).\label{eq:loss}
\end{equation}

We further elaborate on the input features. Industrial recommendation systems rely on a large number of features, which can be categorized into three groups: user features, item features, and context features. Here, we present some features widely used in the industry:
\begin{itemize}
    \item User features: user ID, age, gender, city, operating system, App version, activity level, ID sequence of `Like, Finish, Comment, Share, Follow, Click' behavior and so on.
    \item Item features: item ID, author ID, description, tag, genre, theme and so on.
    \item Context features: time, scenario and so on.
\end{itemize}

There are both ID-based personalized features and non-ID meta features. However, most existing methods~\cite{pan2019warm,ouyang2021learning,zhu2021learning,tang2018personalized,zhou2018deep,pi2020search,chang2023twin} work on the ID-based personalized features, ignoring the non-ID meta features. This paper focuses on comprehensive learning and responsive leveraging all features.

\begin{figure}[t]
	\centering
	\begin{minipage}[b]{0.95\linewidth}
		\centering
		\includegraphics[width=1.\linewidth]{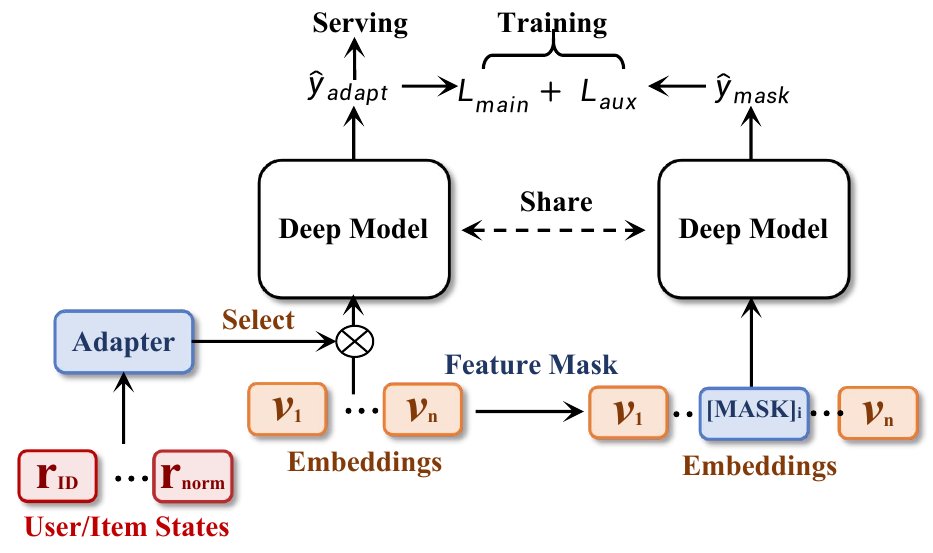}
	\end{minipage}
	\caption{Adaptive Feature Modeling with Feature Mask Framework (AdaF$^2$M$^2$).}\label{fig:model}
\end{figure}

\subsection{Feature Mask}\label{sec:fm}
We propose a feature-mask mechanism to enhance representations via multi-forward task-oriented training with augmented samples, which enables more comprehensive feature learning. The key idea is to get rid of the over-reliance on important features by randomly masking part of all features. We create multiple augmented samples by feature-mask and train on these samples with a task-oriented loss. The augmented samples can force the model to make predictions based on diverse combinations of features, which enables all features to be learned well. 

First, for each instance, we sample $k$ times randomly from a range $[\beta, \gamma]$ and obtain $k$ values, where $\beta$ and $\gamma$ represent a range of sampling probability (setting as 0.1 and 0.5 in this paper). Each value $p$ indicates the probability of replacing a feature embedding $\bm{v}$ with a default mask embedding $[MASK]$. Note that the default mask embedding of each feature is different.
Given the masking probability $p$, we randomly mask some feature embeddings of a sample with corresponding default mask embeddings to generate an augmented sample. The feature embeddings of an augmented sample are formulated as:
\begin{equation}
    [\bm{v}_1, \cdots, [MASK]_i, \cdots, [MASK]_j, \cdots, \bm{v}_n],\label{eq:mask}
\end{equation}
where $[MASK]_i$ indicates the default embedding of the $i$-th feature. With $k$ times random masking, $k$ augmented samples are generated from a given original sample.

The second step is learning with the augmented samples. Inspired by recent natural language processing (NLP)~\cite{devlin2019bert,liu2019roberta} and computer vision (CV)~\cite{chen2020simple,he2020momentum} techniques, most existing recommendation methods~\cite{zhou2020s3,yao2021self} also use self-supervised learning for the augmented samples. The main idea of these methods is that let augmented data from the same sample be discriminated against others, which can improve the discriminability of the representations of users/items. However, the recommendation tasks are different from NLP and CV in two ways: (1) Most tasks of CV and NLP lack sufficient labeled samples, but there are a large amount of unlabeled samples in the real world. However, in recommendation systems, there are billions of labeled samples every day (the feedbacks of users are utilized as labels), and the number of labeled samples is more than the unlabeled samples (the items with no interaction). (2) The prerequisite for many tasks of CV and NLP is understanding the content (text and image), so they need to differentiate between different samples/contents. However, the goal of recommendation models is to make accurate predictions, rather than distinguish different users/items. 

Along this line, we propose multi-forward training with a task-oriented optimization procedure to learn the augmented samples. In detail, the concatenated embeddings of an augmented sample as Equation~(\ref{eq:mask}) are fed into the deep network $g(\cdot)$ to generate the predicted result $\hat{y}$, formulated as:
\begin{equation}
    \hat{y}_{mask} = g([\bm{v}_1, \cdots, [MASK]_i, \cdots, \bm{v}_n]).\label{eq:mask_predict}
\end{equation}

Then, we feed $k$ augmented samples into deep networks to obtain $k$ prediction $\hat{y}_{mask}$.
A task-oriented optimization procedure (directly computing the target loss as Equation~(\ref{eq:loss})) is adopted, which applies additional supervision over the prediction of the augmented samples, denoted as:
\begin{equation}
    \mathcal{L}_{aux} = \sum_{(\bm{x}, y) \in \mathcal{D}} \sum_{i=1}^k -y\log(\hat{y}^i_{mask}) - (1 - y) \log(1 - \hat{y}^i_{mask}),\label{eq:aux}
\end{equation}
where $\mathcal{D}$ indicates the training dataset and $\hat{y}^i_{mask}$ indicates the prediction of the $i$-th random augmented sample. Then, we apply stochastic gradient descent to update all parameters of the deep model and all embeddings.

The feature-mask mechanism which generates multiple augmented samples can help comprehensive feature learning via multi-forward training with task-oriented optimization. It has two main advantages: (1) In real-world recommendation systems, it is common that some features of a sample are missing or mislabeled. The augmented samples are used to simulate the noisy condition of the real-world recommendation systems, which can improve the robustness and generalizability of the model. (2) With the feature-mask mechanism, any features are possible to be masked, including ID-based personalized features. Feeding inputs without personalized features to the deep network can force the model to pay attention to the remained non-ID meta features, e.g., age and gender. Since we adopt task-oriented optimization, the model directly makes predictions based on these unmasked features, which can learn them better. The underlying idea is similar to counterfactual learning, and the counterfactual assumption is that if a high-active user only has the basic features that are assigned in his new-user stage, the model would have consistent prediction for the same user in different stages.

\subsection{Adaptive Feature Modeling}\label{sec:3.3}
In this section, we propose adaptive feature modeling with a state-aware adapter, which can assign adaptive feature weights for users and items with different states. The deep model takes all feature embeddings $[\bm{v}_1, \cdots, \bm{v}_n]$ as input, and the  adaptive feature modeling aims to assign adaptive feature weights, and the input with weights can be formulated as:
\begin{equation}
\begin{split}
    & [a(\bm{v}_1), \cdots, a(\bm{v}_n)], \\
    \text{where}\ \ & a(\bm{v}_i) = w_i \bm{v}_i,
\end{split}
\end{equation}
where $w_i$ indicates the adaptive weight for the $i$-th feature. 

How to generate adaptive weights is the most important key of adaptive feature modeling. The existing methods~\cite{xiao2017attentional,huang2019fibinet,song2019autoint,zhang2021multi,cheng2020adaptive} mainly generate the weights from the feature themselves, which could be formulated as:
\begin{equation}
    [w_1, \cdots, w_n] = h([\bm{v}_1, \cdots, \bm{v}_n]),
\end{equation}
where the $h(\cdot)$ represents the weight generator. However, the learning process is dominated by the head users/items, and the generator would assign bigger weights for the important features of the head users and items. Generally, ID-based personalized features, e.g., ID and sequential features, are important in these methods. 

To tackle the problem of over-reliance on ID-based features in existing methods, we propose a state-aware adapter, which can assign adaptive weights to features according to the states of users/items. To enable the adapter to perceive the different states, we propose four kinds of empirical state signals:


\begin{itemize}
    \item \textbf{Active days}: The number of days users have opened App in the last 7/30 days is used as active-day features, which can distinguish low-, middle-, and high-active users, denoted as $\bm{r}_{active}$. 
    \item \textbf{ID embedding}: With ID embeddings which contain sufficient personalized information as input, the adapter can keep the ability to make personalized predictions for head users and items. We concatenate all ID embedding (user, item, and artist ID) into a vector, denoted as $\bm{r}_{ID}$. 
    \item \textbf{Norm of ID embedding}: Generally, the norm of ID embedding can indicate the quality of ID embedding, e.g., the norm of old users' ID embeddings is bigger than the new users'. Thus, the norm of ID embedding can be exploited to distinguish new and old users, hot and long-tail items. We concatenate the norm of various ID embeddings and use different non-linear functions (log, sqrt, square) to enhance the representations of the norm, denoted as $\bm{r}_{norm}$.  
    \item \textbf{Interaction count}: The number of interactions is a strong signal to distinguish the states of users and items. For example, the low-active users (long-tail items) have less interactions than high-active users (hot items), so a small count indicates low-active users (long-tail items). Thus, we utilize the numbers of `impression', `add comment', `click like button' of users and items, indicated as $\bm{r}_{count}$.
\end{itemize}

Thus, the state-aware adapter takes the concatenated state embeddings $[\bm{r}_{active}, \bm{r}_{ID}, \bm{r}_{norm}, \bm{r}_{count}]$ as input to assign adaptive weight for users/items with different states, formulated as:
\begin{equation}
    [w_1, \cdots, w_n] = \sigma(h([\bm{r}_{active}, \bm{r}_{ID}, \bm{r}_{norm}, \bm{r}_{count}])),
\end{equation}
where $\sigma(\cdot)$ indicates the Sigmoid function. Online experiments show that the performance with the Sigmoid function is better than the Softmax function. The main reason would be that the output of the Softmax function can be dominated by the main features of head users/items. With the guide of the strong empirical signals, the adapter has better ability to generate suitable distributions of weights for different states of users and items. 

In addition, the adaptive mechanism in existing methods introduces another issue: it can seriously influence the learning process of feature embeddings. In detail, the prediction with adaptive weights can be formulated as:
\begin{equation}
    \hat{y} = g([a(\bm{v}_1), \cdots, a(\bm{v}_n)]).
\end{equation}

By computing the gradient and performing a gradient descent step, we obtain a new embedding parameter:
\begin{equation}
\begin{split}
    \bm{v}' = \bm{v} - \lambda \frac{\partial \mathcal{L}}{\partial \bm{v}}
    & = \bm{v} - \lambda \frac{\partial \mathcal{L}}{\partial g}\frac{\partial g}{\partial a}\frac{\partial a}{\partial \bm{v}}, \\
    & = \bm{v} - \lambda w \frac{\partial \mathcal{L}}{\partial g}\frac{\partial g}{\partial a},\label{eq:grad}
\end{split}
\end{equation}
where $\lambda$ indicates the learning rate. Equation~(\ref{eq:grad}) shows the gradient of embedding $\bm{v}$ is related to the feature weight $w$. In other words, a feature assigned with a low weight would be learned with a small gradient (vanishing gradients).
The non-ID meta features which are important for long-tail users/items could be assigned with low weights for high-active users that accounts for most samples. Thus, it is more difficult to learn these features from the samples of head users/items, while the long-tail users/items only accounts a small number of samples.

To alleviate the problem of vanishing gradients in existing methods, we train the state-aware adapter and augmented samples generated by the feature-mask mechanism respectively. In detail, the adaptive weights have no influence on the multi-forward training with the task-oriented optimization, so the feature-mask mechanism can guarantee the comprehensive feature representation learning. Meanwhile, the adapter only works on adaptive feature leveraging for users/items with different states.


\begin{algorithm} [t] 
    \caption{Adaptive Feature Modeling with Feature Mask}\label{alg}
    \flushleft{\hspace*{0.02in}\textbf{Input}: feature embeddings $[\bm{v}_1, \cdots, \bm{v}_n]$, state embeddings $[\bm{r}_{active}, \bm{r}_{ID}, \bm{r}_{norm}, \bm{r}_{count}]$, and the label $y$.\\
    \hspace*{0.02in}\textbf{Input}: The base model $g(\cdot)$ and the state-aware adapter $h(\cdot)$.\\
    \hspace*{0.02in}\textbf{Training Stage}:\\
    \hspace*{0.06in}1. Generate $k$ augmented samples for each sample:\\
    \hspace*{0.46in}$[\bm{v}_1, \cdots, [MASK]_i, \cdots, \bm{v}_n]$. \\
    \hspace*{0.06in}2. Evaluate the task-oriented auxiliary loss $\mathcal{L}_{aux}$ as Equation~(\ref{eq:aux}).\\
    \hspace*{0.06in}3. Compute the state-aware prediction: \\
    \hspace*{0.46in}$\hat{y}_{adapt} = g([a(\bm{v}_1), \cdots, a(\bm{v}_n)])$. \\
    \hspace*{0.06in}4. Evaluate the main loss $\mathcal{L}_{main}$ as Equation~(\ref{eq:adapt_loss}).\\
    \hspace*{0.06in}5. Evaluate the overall loss $\mathcal{L} = \mathcal{L}_{main} + \alpha \mathcal{L}_{aux}$.\\
    \hspace*{0.06in}6. Update all parameters with $\mathcal{L}$.\\
    \hspace*{0.02in}\textbf{Serving Stage}: \\
    \hspace*{0.06in}7. Serve with the state-aware prediction $\hat{y}_{adapt}$.\\
    }
\end{algorithm}

\subsection{Overall Framework}\label{sec:3.4}
The overall framework of AdaF$^2$M$^2$ is shown in Figure~\ref{fig:model}, which consists a state-aware adapter and a feature-mask mechanism. The deployment of the AdaF$^2$M$^2$ framework consists
of two stages: the training stage and the serving stage.

\textbf{Training stage}: The final probability prediction with the weights generated by the state-aware adapter is formulated as:
\begin{equation}
    \hat{y}_{adapt} = g([a(\bm{v}_1), \cdots, a(\bm{v}_n)]),\label{eq:adapt_pred}
\end{equation}
where the $g(\cdot)$ indicates the prediction function. We train the overall framework, including embeddings, the deep network, the adapter, with the cross-entropy loss:
\begin{equation}
    \mathcal{L}_{main} = \sum_{(\bm{x}, y) \in \mathcal{D}} -y\log(\hat{y}_{adapt}) - (1 - y) \log(1 - \hat{y}_{adapt}),\label{eq:adapt_loss}
\end{equation}
where $\mathcal{D}$ indicates the training dataset. For comprehensive feature learning, we utilize the auxiliary loss as shown in Equation~(\ref{eq:aux}). Note that the adaptive weights are not one of the input of the Equation~(\ref{eq:mask_predict}). The overall loss function can be formulated as:



\begin{equation}
    \mathcal{L} = \mathcal{L}_{main} + \alpha \mathcal{L}_{aux}, 
\end{equation}
where the $\alpha$ denotes a hyper-parameter, which is set as 0.2 in this paper. With the training procedure, AdaF$^2$M$^2$ can comprehensively learn features and apply adaptive feature weights responsively to different user/item states.


\textbf{Serving stage}: The serving stage aims at delivering attractive items for the right users, and the model needs to predict the probability that the user might click/like/share each item.  
$\hat{y}_{adapt}$ is directly used as the final prediction in the serving stage. Note that the additional forwards with randomly masked features as input are only utilized in the training stage. Thus, the feature-mask mechanism has no impact on the latency of the serving stage.


\textbf{Deployment}: The proposed AdaF$^2$M$^2$ as shown in Figure~\ref{fig:model} can be applied upon various existing recommendation models~\cite{covington2016deep,he2017neural,cheng2020adaptive,wang2021dcn}, and the overall training and serving procedure is summarized in Algorithm~\ref{alg}. For models of ranking tasks, AdaF$^2$M$^2$ can be directly deployed as Figure~\ref{fig:model}. For two-tower models of retrieval tasks, the user tower and the item tower have different corresponding adapters, and the adapter of the user/item tower only takes the user's/item's state embeddings as input.


\section{Experiments}
In this section, we conduct extensive offline and online experiments with the aim of answering the following evaluation questions: 
\begin{itemize}
     \item[\textbf{EQ1}] Can this AdaF$^2$M$^2$ framework bring improvement to the performance of different online recommendation tasks?
     \item[\textbf{EQ2}] How does the AdaF$^2$M$^2$ framework perform in public and industrial datasets upon various deep recommendation models?
     \item[\textbf{EQ3}] What are the effects of the feature-mask mechanism and the state-aware adapter in our proposed AdaF$^2$M$^2$?
\end{itemize}

\begin{table*}[htbp]
\centering
\setlength\tabcolsep{3pt}
   \caption{Online A/B testing results of the ranking task. Each row indicates the relative improvement with our AdaF$^2$M$^2$ framework over the baseline (a DCN-V2 based multi-task model).  The square brackets represent the 95\% confidence intervals for online metrics. Statistically-significant improvement is marked with bold font in the table.}
\begin{tabular}{lcccccc}
\toprule
\multirow{2}{*}{State}     & \multicolumn{2}{c}{Main Metrics}                    & \multicolumn{4}{c}{Constraint Metrics}                                                                    \\
\cmidrule(r){2-3}  \cmidrule(r){4-7}
                           & Active Days               & Duration                 & Like                     & Finish                   & Comment                  & Dislike                  \\
\midrule
\multirow{2}{*}{New users} & \textbf{+0.333\%}         & \textbf{+0.540\%}          & \textbf{+0.562\%}         & \textbf{+0.418\%}         & \textbf{+0.402\%}         & -2.246\%                 \\
                           & {[}-0.179\%, +0.180\%{]} & {[}-0.370\%, +0.371\%{]} & {[}-0.298\%,+0.297\%{]}  & {[}-0.223\%, +0.223\%{]} & {[}-0.369\%, +0.369\%{]} & {[}-2.984\%, +2.984\%{]} \\
\multirow{2}{*}{Old users} & \textbf{+0.206\%}         & \textbf{+0.434\%}         & \textbf{+0.387\%}         & \textbf{+0.157\%}         & \textbf{+0.230\%}         & -1.394\%                 \\
                           & {[}-0.065\%, +0.065\%{]}   & {[}-0.103\%, +0.104\%{]} & {[}-0.138\%, +0.138\%{]} & {[}-0.084\%, +0.084\%{]} & {[}-0.209\%, +0.208\%{]} & {[}-1.794\%, +1.794\%{]} \\
\midrule
\multirow{2}{*}{Overall}   & \textbf{+0.212\%}         & \textbf{+0.442\%}         & \textbf{+0.418\%}         & \textbf{+0.224\%}         & \textbf{+0.262\%}         & \textbf{-1.594\%}        \\
                           & {[}-0.063\%, +0.063\%{]} & {[}-0.100\%, +0.101\%{]} & {[}-0.126\%, +0.126\%{]} & {[}-0.068\%, +0.068\%{]} & {[}-0.184\%, +0.185\%{]} & {[}-1.541\%, +1.541\%{]} \\
\bottomrule
\end{tabular}\label{tab:online_ranking}
\end{table*}

\begin{table}[htbp]
\caption{Online A/B testing results of the ranking task. We divide old users into low-, middle-, and high-active users to further analyze the effectiveness of AdaF$^2$M$^2$ for users of different states.}
\begin{tabular}{lcc}
\toprule
\multirow{2}{*}{State}               & \multicolumn{2}{c}{Main Metrics}                    \\
\cmidrule(r){2-3}
                                     & Active Days               & Duration                 \\
\midrule
\multirow{2}{*}{Low-active users}    & +0.132\%        & \textbf{+0.356\%}         \\
                                     & {[}-0.229\%, +0.229\%{]} & {[}-0.237\%, +0.237\%{]} \\
\multirow{2}{*}{Middle-active users} & \textbf{+0.195\%}         & \textbf{+0.430\%}         \\
                                     & {[}-0.133\%, +0.133\%{]} & {[}-0.131\%, +0.131\%{]} \\
\multirow{2}{*}{High-active users}   & \textbf{+0.184\%}         & \textbf{+0.425\%}         \\
                                     & {[}-0.080\%, +0.081\%{]}   & {[}-0.112\%, +0.112\%{]}\\
\bottomrule
\end{tabular}\label{tab:online_ranking2}
\end{table}

\subsection{Experimental Settings}\label{sec:exp}
\textbf{Datasets.} We evaluate AdaF$^2$M$^2$ with baselines on both public and large-scale industrial recommendation datasets. 

\textit{DouyinMusic}: Douyin provides a music recommendation service, with over 10 million daily active users. We collect from the impression logs and get two datasets with different sizes, which helps verify the impact of the volume of the dataset. Besides, there is more than a month gap between the two sampled datasets to test the effectiveness of the proposed framework in different periods. The small dataset contains more than 4 billion samples, denoted as \textit{DouyinMusic-4B}. The large dataset contains more than 20 billion samples, denoted as \textit{DouyinMusic-20B}. Each sample of the industrial datasets contains more than one hundred features, including both non-ID meta features (gender, age, genre, mood, scene, and so on) and ID-based personalized features (user ID, item ID, artist ID, interacted ID sequence), which can represent the real-world scenarios. We use `Finish' as the label.

The DouyinMusic-4B dataset contains the recommendation samples from Douyin Music across the time span of 2 weeks in November 2023. Then, we sort all users' prediction events according to timestamp order, taking the first 10 days as the training set, the following 2 days as the validation set, and the remaining 2 days as the test set. The DouyinMusic-20B dataset contains samples from Douyin Music across the time span of 8 weeks from August to September in 2023. Then, we take the first 6 weeks as the training set, the following 1 week as the validation set, and the remaining 1 week as the test set. 

\textit{MovieLens-1M\footnote{http://www.grouplens.org/datasets/movielens/}}: It is one of the most well-known benchmark datasets. The data consists of 1 million movie ranking instances over thousands of movies and users. Each movie has features including its title, year of release, and genres. Titles and genres are lists of tokens. Each user has features including the user’s ID, age, gender, and occupation. We transform ratings into binary (The ratings of at least 4 are turned into 1 and the others are turned into 0). The dataset is randomly divided into train / validation / test sets in the ratio of 6:2:2.

\textbf{Baselines.}
We categorize our baselines into two groups according to their approaches. The first group includes the popular retrieval methods, including:
\begin{itemize}
    \item \textit{Factorization Machines(FM)}~\cite{rendle2010factorization}: FM can capture high-order interaction information. In the industry, three-field FM is widely adopted in retrieval tasks. In detail, user features, item features, and context features are pooled into one embedding respectively. Then, FM is utilized on the three embeddings.
    \item \textit{YouTube DNN}~\cite{covington2016deep}: It is a classical two-tower model, which encodes user features and item features with a deep network respectively. Recently, based on YouTube DNN, many two-tower models~\cite{liu2019real,hao2021adversarial} are proposed.
\end{itemize}

\begin{table}[htbp]
\centering
\setlength\tabcolsep{3pt}
\caption{Online A/B testing results of the retrieval task. We use FM and a two-tower model as base models and show the relative online improvement.}
\begin{tabular}{lcc}
\toprule
\multirow{2}{*}{Method}                                                                           & \multicolumn{2}{c}{Main Metrics}                    \\
\cmidrule(r){2-3}
& Active Days               & Duration                 \\
\midrule
\multirow{2}{*}{\begin{tabular}[c]{@{}l@{}}FM w/ AdaF$^2$M$^2$ \\ \ \ \ \ \ \ v.s. FM\end{tabular}}                   & \textbf{+0.066\%}         & \textbf{+0.195\%}         \\
                                                                                                  & {[}-0.060\%, +0.061\%{]}   & {[}-0.153\%, +0.153\%{]} \\
\multirow{2}{*}{\begin{tabular}[c]{@{}l@{}}Two Tower w/ AdaF$^2$M$^2$ \\ \ \ \ \ \ \ v.s. Two Tower\end{tabular}} & \textbf{+0.073\%}         & \textbf{+0.184\%}         \\
                                                                                                  & {[}-0.046\%, +0.046\%{]} & {[}-0.086\%, +0.086\%{]}\\
\bottomrule
\end{tabular}\label{tab:matching}
\end{table}

The second group includes state-of-the-art methods for ranking tasks in recommendation systems, including:
\begin{itemize}
    \item \textit{Adaptive Factorization Network(AFN)}~\cite{cheng2020adaptive}: It is a recent FM-based~\cite{rendle2010factorization} method which learns arbitrary-order cross features adaptively from data. The core of AFN is a logarithmic transformation layer that converts the power of each feature in a feature combination into the learnable coefficient.
    \item \textit{Deep \& Cross Network V2(DCN-V2)}~\cite{wang2021dcn}: In light of the pros/cons of DCN and existing feature interaction learning approaches, DCN-V2 is an improved framework to make DCN~\cite{wang2017deep} more practical in large-scale industrial settings, which is more expressive yet remains cost efficient at feature interaction learning, especially when coupled with a mixture of low-rank architecture.
    \item \textit{EulerNet}~\cite{tian2023eulernet}: It is a recent state-of-the-art method, in which the feature interactions are learned in a complex vector space by conducting space mapping according to Euler’s formula. Furthermore, EulerNet incorporates the implicit and explicit feature interactions into a unified architecture.
\end{itemize}

The proposed AdaF$^2$M$^2$ is a model-agnostic framework that can be applied upon various models, and we apply AdaF$^2$M$^2$ on the above five base models to demonstrate its effectiveness and universality.

\subsection{Online A/B Testing (EQ1)} 

To verify the real benefits AdaF$^2$M$^2$ brings to our system, we conducted online A/B testing experiments for more than two weeks for the ranking, retrieval, and item cold-start tasks in Douyin Music App respectively. Indeed, the proposed AdaF$^2$M$^2$ has been deployed to the online ranking, retrieval, and item cold-start tasks in multiple applications of Douyin Group.

\begin{table}[tbp]
\centering
\caption{Online A/B testing results of the item cold-start tasks. We divide items into multiple groups according to impression counts (Item Group). The impr\_num and item\_num indicate the number of impressions and items in the item group.}
\begin{tabular}{lcccc}
\toprule
Item Group                 & Impr\_num & Item\_num & Like     & Finish  \\
\midrule
0-128            & \textbf{+19.638\%}   & \textbf{+35.836\%}   & \textbf{+18.495\%} & \textbf{+0.339\%} \\
128-512 & \textbf{+8.240\%}    & \textbf{+1.892\%}    & \textbf{+3.153\%}  & \textbf{+2.917\%} \\
512-1024         & \textbf{+8.701\%}    & +0.128\%    & \textbf{+4.220\%}  & \textbf{+3.820\%}\\
\bottomrule
\end{tabular}\label{tab:cold}
\end{table}

\begin{table}[tbp]
\centering
\caption{Offline results (RelaImpr of AUC and UAUC) on the industrial datasets DouyinMusic-20B and DouyinMusic-4B.}
\begin{tabular}{lcccc}
\toprule
            & \multicolumn{2}{c}{DouyinMusic-20B} & \multicolumn{2}{c}{DouyinMusic-4B} \\
\cmidrule(r){2-3} \cmidrule(r){4-5}
Method      & AUC              & UAUC             & AUC              & UAUC            \\
\midrule
FM          & \textbf{+0.41\%}  & \textbf{+0.26\%}  & \textbf{+0.46\%}  & \textbf{+0.37\%} \\
YouTube DNN & \textbf{+0.15\%}  & \textbf{+0.17\%}  & \textbf{+0.18\%}  & \textbf{+0.22\%} \\
AFN         & \textbf{+0.25\%}  & \textbf{+0.13\%}  & \textbf{+0.23\%}  & \textbf{+0.15\%} \\
DCN-V2      & \textbf{+0.21\%}  & \textbf{+0.15\%}  & \textbf{+0.45\%}  & \textbf{+0.71\%} \\
EulerNet    & \textbf{+0.19\%}  & \textbf{+0.17\%}  & \textbf{+0.36\%}  & \textbf{+0.43\%} \\
\bottomrule
\end{tabular}\label{tab:offline1}
\end{table}

\textbf{Online Metrics}. We evaluate model performance based on two main metrics, Active Days and Duration, which are widely adopted in practical recommendation systems. The total days that users in the experimental bucket open the application are denoted as Active Days. The total amount of time spent by users in the experimental bucket on staying in the application is denoted as Duration. We also take additional metrics, which evaluate user engagement, including Like/Dislike (clicking the like/dislike button on the screen), Finish (hearing the end of a song), and Comment (leaving comments on a song), which are usually used as constraint metrics. We calculate all online metrics per user.

\textbf{Ranking Tasks}. We apply the proposed AdaF$^2$M$^2$ on a DCN-V2-based multi-task model which is deployed in the online ranking tasks and conduct an online A/B testing to demonstrate the effectiveness of AdaF$^2$M$^2$ to improve ranking models. The online A/B results of new users (the registration time is less than X days from now, and X is a predefined threshold), old users (the registration time is greater than X days), and whole users are shown in Table~\ref{tab:online_ranking}. In addition, the old users are further divided into low-, middle-, and high-active users according to the active days in the recent 30 days, and the results are shown in Table~\ref{tab:online_ranking2}. For the main metrics Active Days and Duration, the proposed AdaF$^2$M$^2$ achieves a large improvement of +0.212\% and +0.442\% for all users with statistical significance, which is remarkable given the fact that the average Active Days and Duration improvement from production algorithms is around 0.05\% and 0.1\%, respectively. In addition, the results demonstrate that AdaF$^2$M$^2$ could improve the recommendation performance for users of different states.

\begin{table}[thbp]
  \centering
  \setlength\tabcolsep{15pt}
   \caption{Offline results (AUC, RelaImpr) on the public dataset MovieLens-1M.}
    \begin{tabular}{lcc}
    \toprule
    Method & AUC & RelaImpr\\
    \midrule
    FM & 0.7767 & -  \\
    \ \ \ \  w/ AdaF$^2$M$^2$ & \textbf{0.7808} & +0.53\% \\
    YouTube DNN & 0.7857 & - \\
    \ \ \ \  w/ AdaF$^2$M$^2$ & \textbf{0.7871} & +0.18\%\\
    AFN & 0.7878 & - \\
    \ \ \ \  w/ AdaF$^2$M$^2$ & \textbf{0.7889} & +0.14\%\\
    DCNV2 & 0.7886 & -\\
    \ \ \ \  w/ AdaF$^2$M$^2$ & \textbf{0.7905} & +0.24\%\\
    EulerNet & 0.7911 & - \\
    \ \ \ \  w/ AdaF$^2$M$^2$ & \textbf{0.7921} & +0.13\%\\
    \bottomrule
    \end{tabular}%
  \label{tab:offline2}%
\end{table}%

\textbf{Retrieval Tasks}. We deploy AdaF$^2$M$^2$ on two online retrieval models, including an FM-based model and a two-tower model that is similar to YouTube DNN~\cite{covington2016deep}. The online A/B results of the retrieval tasks are shown in Table~\ref{tab:matching}. From the results, we find the proposed AdaF$^2$M$^2$ framework achieves a significant improvement on 
Active Days and Duration, against the two base retrieval models. Especially, an upgrade of a retrieval model can increase about 0.07\% Active Days, which is a very significant improvement. In addition, with different base models, AdaF$^2$M$^2$ is effective in improving the recommendation performance, which demonstrates that AdaF$^2$M$^2$ has satisfying universality.

\textbf{Item Cold-start Tasks}. Practical recommendation systems create a strong feedback loop~\cite{chaney2018algorithmic} resulting in ``rich gets richer'' effect~\cite{wang2023fresh}. Cold-start items face a significant barrier to being picked up by the systems and shown to the right users due to lack of initial exposure and interaction~\cite{wang2023fresh}, while the hot items can be naturally recommended well.  Thus, to demonstrate the effectiveness of AdaF$^2$M$^2$ for items, we apply AdaF$^2$M$^2$ on a two-tower cold-start model and conduct online A/B testing. We divide items into three groups according to the number of impressions, denoted as 0-128, 128-512, and 512-1024, and the results are shown in Table~\ref{tab:cold}. From the results, we find that AdaF$^2$M$^2$ can improve the number of impressions, items, Like, and Finish for cold-start items, which demonstrates AdaF$^2$M$^2$ is able to identify more quality cold-start items that can attract user interactions.

\subsection{Offline Experiments (EQ2)}
In this section, we conduct extended offline experiments with one public dataset and two industrial datasets. Specifically, we apply the proposed AdaF$^2$M$^2$ framework upon five base models to testify the effectiveness and universality.

\textbf{Experimental Details.} For all methods, the initial learning rate for the Adam~\cite{kingma2014adam} optimizer are tuned by grid searches within \{0.001, 0.005, 0.01, 0.02, 0.1\}. In addition, we tune the dimension of embeddings in \{16, 32, 64, 128, 256\}. For all methods, we set a mini-batch size of 256. For all MLPs in these methods, the ReLU activation function is employed, and the dimension of each layer is set to 512, 128, and 1. $k$ is set as 1. We report the offline results via three random runs. Statistically significant improvement is marked with bold font in the tables.

\begin{table}[th]
\centering
\setlength\tabcolsep{3pt}
\caption{Online ablation experiments of the retrieval tasks. We use a two-tower model as the base model and show the relative online improvement.}
\begin{tabular}{lcc}
\toprule
\multirow{2}{*}{Method}          & \multicolumn{2}{c}{Main Metrics}                    \\
\cmidrule(r){2-3}
                                 & Active Days               & Duration                 \\
\midrule
\multirow{2}{*}{w/ Feature Mask} & +0.006\%         & +0.078\%         \\
                                 & {[}-0.045\%, +0.045\%{]} & {[}-0.125\%, +0.125\%{]} \\
\multirow{2}{*}{w/ Adapter}      & +0.022\%         & +0.091\%         \\
                                 & {[}-0.063\%, +0.063\%{]} & {[}-0.153\%, +0.153\%{]} \\
\multirow{2}{*}{w/ AdaF$^2$M$^2$}        & \textbf{+0.073\%}         & \textbf{+0.184\%}         \\
                                 & {[}-0.046\%, +0.046\%{]} & {[}-0.086\%, +0.086\%{]} \\
\midrule
\multirow{2}{*}{w/ MaskNet}  & +0.033\%                  & +0.048\%                  \\
                                 & {[}-0.073\%, +0.073\%{]} & {[}-0.244\%, +0.244\%{]}\\
\bottomrule
\end{tabular}\label{tab:ablation1}
\end{table}

\begin{table}[th]
\centering
\setlength\tabcolsep{15pt}
\caption{Ablation study with a two-tower model as the base model on the DouyinMusic-20B dataset.}
\begin{tabular}{lcc}
\toprule
\multicolumn{1}{c}{Method} & AUC & UAUC           \\
\midrule
w/ Feature Mask      & +0.05\%          & +0.03\%          \\
w/ Adapter           & \textbf{+0.12\%} & \textbf{+0.11\%} \\
w/ AdaF$^2$M$^2$            & \textbf{+0.15\%}  & \textbf{+0.17\%} \\
\midrule
w/ MaskNet & \textbf{+0.14\%} & \textbf{+0.12\%} \\
\bottomrule
\end{tabular}\label{tab:ablation2}
\end{table}

\textbf{Offline Metrics.} For binary classification tasks, AUC is a widely used metric~\cite{fawcett2006introduction}. It measures the goodness of order by ranking all the items with prediction, including intra-user and inter-user orders. Besides, AUC is a common metric for recommendation~\cite{he2017neural,zhou2018deep}. In addition, we introduce another metric in our business, named UAUC, which calculates AUC per user and then averages scores with weights proportional to the user's sample size. A larger UAUC suggests better model performance. For the industrial datasets, we report the relative improvement (RelaImpr) of AUC and UAUC over base models (the limitation of our company). The public dataset only contains 6,000 users and the user-level AUC(UAUC) is not very confident. Thus, following most existing methods~\cite{tian2023eulernet,he2017neural}, we report AUC and RelaImpr.

\textbf{Offline Results.} The experimental results on industrial and public datasets are shown in Table~\ref{tab:offline1} and Table~\ref{tab:offline2}, respectively. The results further reveal several insightful observations. With the help of the proposed AdaF$^2$M$^2$, most base models show a significant improvement, which demonstrates the effectiveness and universality of the proposed framework. The improvement mainly comes from the better feature representation learning and leveraging mechanisms. In addition, it also testifies that AdaF$^2$M$^2$ is a model-agnostic framework that can be applied upon various base models. 

\begin{figure}[t]
	\centering
	\begin{minipage}[b]{1.0\linewidth}
        \subfigure[Analysis of user features]{
            \centering
            \includegraphics[width=1.0\columnwidth]{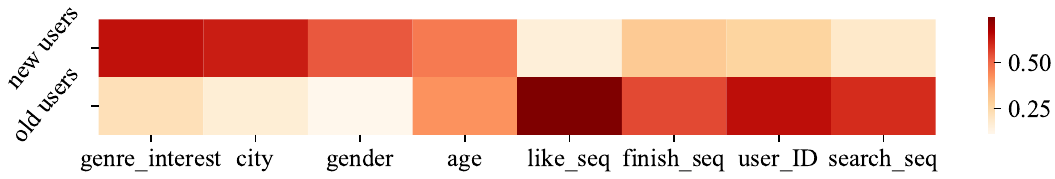}
            \label{fig:analysis_user}
        }
        \subfigure[Analysis of item features]{
            \centering
            \includegraphics[width=1.0\columnwidth]{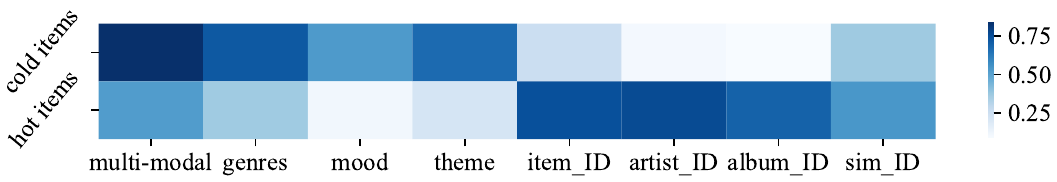}
            \label{fig:analysis_item}
        }
	\end{minipage}
        \vspace{-0.2cm}
	\caption{Heatmaps with the average feature weights of users/items of different states.}\label{fig:analysis}
 \vspace{-0.2cm}
\end{figure}

\subsection{Analysis (EQ3)}
In this section, we analyze the effects of the adapter and the feature-mask mechanism in our proposed AdaF$^2$M$^2$. First, we conduct an ablation study to verify the effectiveness of each module. Second, we analyze the important features of users/items of different states.

\textbf{Ablation Study}. To test the effectiveness of each module in AdaF$^2$M$^2$, we present an ablation study. We introduce three kinds of models, w/ Feature Mask, w/ Adapter, and
w/ AdaF$^2$M$^2$, which add the feature-mask mechanism, the adapter, and the overall AdaF$^2$M$^2$ based on a two-tower model, respectively. The online and offline results are shown in Table~\ref{tab:ablation1}. The results show that the base model with AdaF$^2$M$^2$ achieves better performance than the model only with Feature Mask or Adapter, which demonstrates each module is effective. In addition, we find that the model with the adapter can improve the offline performance, but there is no improvement in the online results. The main reason could be the gap between the offline and online metrics. The offline metrics (AUC and UAUC) are dominated by the performance of the head users, while the online metrics average scores with the same weights for all users. In addition, to demonstrate the proposed AdaF$^2$M$^2$ is better than existing mask methods with self-supervised learning (SSL), we conduct another online A/B testing that replaces the AdaF$^2$M$^2$ with MaskNet~\cite{yao2021self}. The results show that AdaF$^2$M$^2$ is better than MaskNet with the self-supervised loss in recommendation systems.

\textbf{The Effectiveness of Adapter}. To demonstrate the effectiveness of the adapter, we visualize the importance of several typical features (the limitation of pages) in Figure~\ref{fig:analysis}. We find that \{genre\_interest, city, gender, age\} are important for new users, while \{like sequence, finish sequence, search sequence, user ID\} are important for old users. Besides, the predictive features for cold items and hot items are 
\{multi-modal features, genres, mood, theme\} and \{item ID, artist ID, album ID, similar items' ID\} respectively.
The results of the analysis are consistent with practical experience, which demonstrates that AdaF$^2$M$^2$ could capture different distributions of feature importance for different user/item states.

\section{Conclusion}
In this paper, for better feature representation learning and leveraging in recommendation systems, we propose Adaptive Feature Modeling with Feature Mask (AdaF$^2$M$^2$), a novel framework consisting of feature-mask mechanism and a state-aware adapter. 
AdaF$^2$M$^2$ can comprehensively learn features and adaptively leverage features responsive to different user/item states.
We demonstrated the superior performance of the proposed AdaF$^2$M$^2$ in offline experiments. In addition, we conducted online A/B testing in ranking, retrieval, and item cold-start tasks 
on multiple recommendation scenarios, obtaining +1.37\% and +1.89\% cumulative improvements on user active days and app duration respectively, which demonstrates the effectiveness and universality of AdaF$^2$M$^2$ in online systems. 
Moreover, AdaF$^2$M$^2$ has been deployed on both ranking and retrieval tasks in multiple applications of Douyin Group.

\bibliographystyle{ACM-Reference-Format}
\bibliography{sample-base}


\begin{thebibliography}{60}


\ifx \showCODEN    \undefined \def \showCODEN     #1{\unskip}     \fi
\ifx \showDOI      \undefined \def \showDOI       #1{#1}\fi
\ifx \showISBNx    \undefined \def \showISBNx     #1{\unskip}     \fi
\ifx \showISBNxiii \undefined \def \showISBNxiii  #1{\unskip}     \fi
\ifx \showISSN     \undefined \def \showISSN      #1{\unskip}     \fi
\ifx \showLCCN     \undefined \def \showLCCN      #1{\unskip}     \fi
\ifx \shownote     \undefined \def \shownote      #1{#1}          \fi
\ifx \showarticletitle \undefined \def \showarticletitle #1{#1}   \fi
\ifx \showURL      \undefined \def \showURL       {\relax}        \fi
\providecommand\bibfield[2]{#2}
\providecommand\bibinfo[2]{#2}
\providecommand\natexlab[1]{#1}
\providecommand\showeprint[2][]{arXiv:#2}

\bibitem[Bhagat et~al\mbox{.}(2018)]%
        {bhagat2018buy}
\bibfield{author}{\bibinfo{person}{Rahul Bhagat}, \bibinfo{person}{Srevatsan Muralidharan}, \bibinfo{person}{Alex Lobzhanidze}, {and} \bibinfo{person}{Shankar Vishwanath}.} \bibinfo{year}{2018}\natexlab{}.
\newblock \showarticletitle{Buy it again: Modeling repeat purchase recommendations}. In \bibinfo{booktitle}{\emph{Proceedings of the 24th ACM SIGKDD International Conference on Knowledge Discovery \& Data Mining}}. \bibinfo{pages}{62--70}.
\newblock


\bibitem[Cao et~al\mbox{.}(2022a)]%
        {cao2022disencdr}
\bibfield{author}{\bibinfo{person}{Jiangxia Cao}, \bibinfo{person}{Xixun Lin}, \bibinfo{person}{Xin Cong}, \bibinfo{person}{Jing Ya}, \bibinfo{person}{Tingwen Liu}, {and} \bibinfo{person}{Bin Wang}.} \bibinfo{year}{2022}\natexlab{a}.
\newblock \showarticletitle{Disencdr: Learning disentangled representations for cross-domain recommendation}. In \bibinfo{booktitle}{\emph{Proceedings of the 45th International ACM SIGIR conference on research and development in information retrieval}}. \bibinfo{pages}{267--277}.
\newblock


\bibitem[Cao et~al\mbox{.}(2022b)]%
        {cao2022cross}
\bibfield{author}{\bibinfo{person}{Jiangxia Cao}, \bibinfo{person}{Jiawei Sheng}, \bibinfo{person}{Xin Cong}, \bibinfo{person}{Tingwen Liu}, {and} \bibinfo{person}{Bin Wang}.} \bibinfo{year}{2022}\natexlab{b}.
\newblock \showarticletitle{Cross-domain recommendation to cold-start users via variational information bottleneck}. In \bibinfo{booktitle}{\emph{2022 IEEE 38th International Conference on Data Engineering (ICDE)}}. IEEE, \bibinfo{pages}{2209--2223}.
\newblock


\bibitem[Chaney et~al\mbox{.}(2018)]%
        {chaney2018algorithmic}
\bibfield{author}{\bibinfo{person}{Allison~JB Chaney}, \bibinfo{person}{Brandon~M Stewart}, {and} \bibinfo{person}{Barbara~E Engelhardt}.} \bibinfo{year}{2018}\natexlab{}.
\newblock \showarticletitle{How algorithmic confounding in recommendation systems increases homogeneity and decreases utility}. In \bibinfo{booktitle}{\emph{Proceedings of the 12th ACM Conference on Recommender Systems}}. \bibinfo{pages}{224--232}.
\newblock


\bibitem[Chang et~al\mbox{.}(2023)]%
        {chang2023twin}
\bibfield{author}{\bibinfo{person}{Jianxin Chang}, \bibinfo{person}{Chenbin Zhang}, \bibinfo{person}{Zhiyi Fu}, \bibinfo{person}{Xiaoxue Zang}, \bibinfo{person}{Lin Guan}, \bibinfo{person}{Jing Lu}, \bibinfo{person}{Yiqun Hui}, \bibinfo{person}{Dewei Leng}, \bibinfo{person}{Yanan Niu}, \bibinfo{person}{Yang Song}, {et~al\mbox{.}}} \bibinfo{year}{2023}\natexlab{}.
\newblock \showarticletitle{TWIN: TWo-stage Interest Network for Lifelong User Behavior Modeling in CTR Prediction at Kuaishou}. In \bibinfo{booktitle}{\emph{Proceedings of the 29th ACM SIGKDD International Conference on Knowledge Discovery \& Data Mining}}.
\newblock


\bibitem[Chen et~al\mbox{.}(2019)]%
        {chen2019follow}
\bibfield{author}{\bibinfo{person}{Jingwu Chen}, \bibinfo{person}{Fuzhen Zhuang}, \bibinfo{person}{Tianxin Wang}, \bibinfo{person}{Leyu Lin}, \bibinfo{person}{Feng Xia}, \bibinfo{person}{Lihuan Du}, {and} \bibinfo{person}{Qing He}.} \bibinfo{year}{2019}\natexlab{}.
\newblock \showarticletitle{Follow the Title Then Read the Article: Click-Guide Network for Dwell Time Prediction}.
\newblock \bibinfo{journal}{\emph{IEEE Transactions on Knowledge and Data Engineering}} \bibinfo{volume}{33}, \bibinfo{number}{7} (\bibinfo{year}{2019}), \bibinfo{pages}{2903--2913}.
\newblock


\bibitem[Chen et~al\mbox{.}(2020)]%
        {chen2020simple}
\bibfield{author}{\bibinfo{person}{Ting Chen}, \bibinfo{person}{Simon Kornblith}, \bibinfo{person}{Mohammad Norouzi}, {and} \bibinfo{person}{Geoffrey Hinton}.} \bibinfo{year}{2020}\natexlab{}.
\newblock \showarticletitle{A simple framework for contrastive learning of visual representations}. In \bibinfo{booktitle}{\emph{Proceedings of the 37th International Conference on Machine Learning}}. \bibinfo{pages}{1597--1607}.
\newblock


\bibitem[Cheng et~al\mbox{.}(2020)]%
        {cheng2020adaptive}
\bibfield{author}{\bibinfo{person}{Weiyu Cheng}, \bibinfo{person}{Yanyan Shen}, {and} \bibinfo{person}{Linpeng Huang}.} \bibinfo{year}{2020}\natexlab{}.
\newblock \showarticletitle{Adaptive factorization network: Learning adaptive-order feature interactions}. In \bibinfo{booktitle}{\emph{Proceedings of the AAAI Conference on Artificial Intelligence}}, Vol.~\bibinfo{volume}{34}. \bibinfo{pages}{3609--3616}.
\newblock


\bibitem[Covington et~al\mbox{.}(2016)]%
        {covington2016deep}
\bibfield{author}{\bibinfo{person}{Paul Covington}, \bibinfo{person}{Jay Adams}, {and} \bibinfo{person}{Emre Sargin}.} \bibinfo{year}{2016}\natexlab{}.
\newblock \showarticletitle{Deep neural networks for youtube recommendations}. In \bibinfo{booktitle}{\emph{Proceedings of the 10th ACM Conference on Recommender Systems}}. \bibinfo{pages}{191--198}.
\newblock


\bibitem[Deng et~al\mbox{.}(2021)]%
        {deng2021deeplight}
\bibfield{author}{\bibinfo{person}{Wei Deng}, \bibinfo{person}{Junwei Pan}, \bibinfo{person}{Tian Zhou}, \bibinfo{person}{Deguang Kong}, \bibinfo{person}{Aaron Flores}, {and} \bibinfo{person}{Guang Lin}.} \bibinfo{year}{2021}\natexlab{}.
\newblock \showarticletitle{Deeplight: Deep lightweight feature interactions for accelerating ctr predictions in ad serving}. In \bibinfo{booktitle}{\emph{Proceedings of the 14th ACM International Conference on Web Search and Data Mining}}. \bibinfo{pages}{922--930}.
\newblock


\bibitem[Devlin et~al\mbox{.}(2019)]%
        {devlin2019bert}
\bibfield{author}{\bibinfo{person}{Jacob Devlin}, \bibinfo{person}{Ming-Wei Chang}, \bibinfo{person}{Kenton Lee}, {and} \bibinfo{person}{Kristina Toutanova}.} \bibinfo{year}{2019}\natexlab{}.
\newblock \showarticletitle{BERT: Pre-training of Deep Bidirectional Transformers for Language Understanding}. In \bibinfo{booktitle}{\emph{Proceedings of the 2019 Conference of the North American Chapter of the Association for Computational Linguistics: Human Language Technologies}}. \bibinfo{pages}{4171--4186}.
\newblock


\bibitem[Fawcett(2006)]%
        {fawcett2006introduction}
\bibfield{author}{\bibinfo{person}{Tom Fawcett}.} \bibinfo{year}{2006}\natexlab{}.
\newblock \showarticletitle{An introduction to ROC analysis}.
\newblock \bibinfo{journal}{\emph{Pattern recognition letters}} \bibinfo{volume}{27}, \bibinfo{number}{8} (\bibinfo{year}{2006}), \bibinfo{pages}{861--874}.
\newblock


\bibitem[Guo et~al\mbox{.}(2017)]%
        {guo2017deepfm}
\bibfield{author}{\bibinfo{person}{Huifeng Guo}, \bibinfo{person}{TANG Ruiming}, \bibinfo{person}{Yunming Ye}, \bibinfo{person}{Zhenguo Li}, {and} \bibinfo{person}{Xiuqiang He}.} \bibinfo{year}{2017}\natexlab{}.
\newblock \showarticletitle{DeepFM: A Factorization-Machine based Neural Network for CTR Prediction}. In \bibinfo{booktitle}{\emph{Proceedings of the Twenty-Sixth International Joint Conference on Artificial Intelligence}}. International Joint Conferences on Artificial Intelligence Organization.
\newblock


\bibitem[Hao et~al\mbox{.}(2021)]%
        {hao2021adversarial}
\bibfield{author}{\bibinfo{person}{Xiaobo Hao}, \bibinfo{person}{Yudan Liu}, \bibinfo{person}{Ruobing Xie}, \bibinfo{person}{Kaikai Ge}, \bibinfo{person}{Linyao Tang}, \bibinfo{person}{Xu Zhang}, {and} \bibinfo{person}{Leyu Lin}.} \bibinfo{year}{2021}\natexlab{}.
\newblock \showarticletitle{Adversarial feature translation for multi-domain recommendation}. In \bibinfo{booktitle}{\emph{Proceedings of the 27th ACM SIGKDD Conference on Knowledge Discovery \& Data Mining}}. \bibinfo{pages}{2964--2973}.
\newblock


\bibitem[He et~al\mbox{.}(2020)]%
        {he2020momentum}
\bibfield{author}{\bibinfo{person}{Kaiming He}, \bibinfo{person}{Haoqi Fan}, \bibinfo{person}{Yuxin Wu}, \bibinfo{person}{Saining Xie}, {and} \bibinfo{person}{Ross Girshick}.} \bibinfo{year}{2020}\natexlab{}.
\newblock \showarticletitle{Momentum contrast for unsupervised visual representation learning}. In \bibinfo{booktitle}{\emph{Proceedings of the IEEE/CVF Conference on Computer Vision and Pattern Recognition}}. \bibinfo{pages}{9729--9738}.
\newblock


\bibitem[He et~al\mbox{.}(2017)]%
        {he2017neural}
\bibfield{author}{\bibinfo{person}{Xiangnan He}, \bibinfo{person}{Lizi Liao}, \bibinfo{person}{Hanwang Zhang}, \bibinfo{person}{Liqiang Nie}, \bibinfo{person}{Xia Hu}, {and} \bibinfo{person}{Tat-Seng Chua}.} \bibinfo{year}{2017}\natexlab{}.
\newblock \showarticletitle{Neural collaborative filtering}. In \bibinfo{booktitle}{\emph{Proceedings of the 26th International Conference on World Wide Web}}. \bibinfo{pages}{173--182}.
\newblock


\bibitem[Huang et~al\mbox{.}(2019)]%
        {huang2019fibinet}
\bibfield{author}{\bibinfo{person}{Tongwen Huang}, \bibinfo{person}{Zhiqi Zhang}, {and} \bibinfo{person}{Junlin Zhang}.} \bibinfo{year}{2019}\natexlab{}.
\newblock \showarticletitle{FiBiNET: combining feature importance and bilinear feature interaction for click-through rate prediction}. In \bibinfo{booktitle}{\emph{Proceedings of the 13th ACM Conference on Recommender Systems}}. \bibinfo{pages}{169--177}.
\newblock


\bibitem[Juan et~al\mbox{.}(2016)]%
        {juan2016field}
\bibfield{author}{\bibinfo{person}{Yuchin Juan}, \bibinfo{person}{Yong Zhuang}, \bibinfo{person}{Wei-Sheng Chin}, {and} \bibinfo{person}{Chih-Jen Lin}.} \bibinfo{year}{2016}\natexlab{}.
\newblock \showarticletitle{Field-aware factorization machines for CTR prediction}. In \bibinfo{booktitle}{\emph{Proceedings of the 10th ACM conference on recommender systems}}. \bibinfo{pages}{43--50}.
\newblock


\bibitem[Kingma and Ba(2015)]%
        {kingma2014adam}
\bibfield{author}{\bibinfo{person}{Diederik~P Kingma} {and} \bibinfo{person}{Jimmy Ba}.} \bibinfo{year}{2015}\natexlab{}.
\newblock \showarticletitle{Adam: A method for stochastic optimization}. In \bibinfo{booktitle}{\emph{International Conference on Learning Representations}}, Vol.~\bibinfo{volume}{5}.
\newblock


\bibitem[Koren et~al\mbox{.}(2009)]%
        {koren2009matrix}
\bibfield{author}{\bibinfo{person}{Yehuda Koren}, \bibinfo{person}{Robert Bell}, {and} \bibinfo{person}{Chris Volinsky}.} \bibinfo{year}{2009}\natexlab{}.
\newblock \showarticletitle{Matrix factorization techniques for recommender systems}.
\newblock \bibinfo{journal}{\emph{Computer}} \bibinfo{volume}{42}, \bibinfo{number}{8} (\bibinfo{year}{2009}), \bibinfo{pages}{30--37}.
\newblock


\bibitem[Lian et~al\mbox{.}(2018)]%
        {lian2018xdeepfm}
\bibfield{author}{\bibinfo{person}{Jianxun Lian}, \bibinfo{person}{Xiaohuan Zhou}, \bibinfo{person}{Fuzheng Zhang}, \bibinfo{person}{Zhongxia Chen}, \bibinfo{person}{Xing Xie}, {and} \bibinfo{person}{Guangzhong Sun}.} \bibinfo{year}{2018}\natexlab{}.
\newblock \showarticletitle{xdeepfm: Combining explicit and implicit feature interactions for recommender systems}. In \bibinfo{booktitle}{\emph{Proceedings of the 24th ACM SIGKDD International Conference on Knowledge Discovery \& Data Mining}}. \bibinfo{pages}{1754--1763}.
\newblock


\bibitem[Liu et~al\mbox{.}(2020a)]%
        {liu2020autogroup}
\bibfield{author}{\bibinfo{person}{Bin Liu}, \bibinfo{person}{Niannan Xue}, \bibinfo{person}{Huifeng Guo}, \bibinfo{person}{Ruiming Tang}, \bibinfo{person}{Stefanos Zafeiriou}, \bibinfo{person}{Xiuqiang He}, {and} \bibinfo{person}{Zhenguo Li}.} \bibinfo{year}{2020}\natexlab{a}.
\newblock \showarticletitle{AutoGroup: Automatic feature grouping for modelling explicit high-order feature interactions in CTR prediction}. In \bibinfo{booktitle}{\emph{Proceedings of the 43rd International ACM SIGIR Conference on Research and Development in Information Retrieval}}. \bibinfo{pages}{199--208}.
\newblock


\bibitem[Liu et~al\mbox{.}(2020b)]%
        {liu2020autofis}
\bibfield{author}{\bibinfo{person}{Bin Liu}, \bibinfo{person}{Chenxu Zhu}, \bibinfo{person}{Guilin Li}, \bibinfo{person}{Weinan Zhang}, \bibinfo{person}{Jincai Lai}, \bibinfo{person}{Ruiming Tang}, \bibinfo{person}{Xiuqiang He}, \bibinfo{person}{Zhenguo Li}, {and} \bibinfo{person}{Yong Yu}.} \bibinfo{year}{2020}\natexlab{b}.
\newblock \showarticletitle{Autofis: Automatic feature interaction selection in factorization models for click-through rate prediction}. In \bibinfo{booktitle}{\emph{proceedings of the 26th ACM SIGKDD International Conference on Knowledge Discovery \& Data Mining}}. \bibinfo{pages}{2636--2645}.
\newblock


\bibitem[Liu et~al\mbox{.}(2019a)]%
        {liu2019real}
\bibfield{author}{\bibinfo{person}{Yudan Liu}, \bibinfo{person}{Kaikai Ge}, \bibinfo{person}{Xu Zhang}, {and} \bibinfo{person}{Leyu Lin}.} \bibinfo{year}{2019}\natexlab{a}.
\newblock \showarticletitle{Real-time attention based look-alike model for recommender system}. In \bibinfo{booktitle}{\emph{Proceedings of the 25th ACM SIGKDD International Conference on Knowledge Discovery \& Data Mining}}. \bibinfo{pages}{2765--2773}.
\newblock


\bibitem[Liu et~al\mbox{.}(2019b)]%
        {liu2019roberta}
\bibfield{author}{\bibinfo{person}{Yinhan Liu}, \bibinfo{person}{Myle Ott}, \bibinfo{person}{Naman Goyal}, \bibinfo{person}{Jingfei Du}, \bibinfo{person}{Mandar Joshi}, \bibinfo{person}{Danqi Chen}, \bibinfo{person}{Omer Levy}, \bibinfo{person}{Mike Lewis}, \bibinfo{person}{Luke Zettlemoyer}, {and} \bibinfo{person}{Veselin Stoyanov}.} \bibinfo{year}{2019}\natexlab{b}.
\newblock \showarticletitle{Roberta: A robustly optimized bert pretraining approach}.
\newblock \bibinfo{journal}{\emph{arXiv preprint arXiv:1907.11692}} (\bibinfo{year}{2019}).
\newblock


\bibitem[Man et~al\mbox{.}(2017)]%
        {man2017cross}
\bibfield{author}{\bibinfo{person}{Tong Man}, \bibinfo{person}{Huawei Shen}, \bibinfo{person}{Xiaolong Jin}, {and} \bibinfo{person}{Xueqi Cheng}.} \bibinfo{year}{2017}\natexlab{}.
\newblock \showarticletitle{Cross-domain recommendation: an embedding and mapping approach}. In \bibinfo{booktitle}{\emph{Proceedings of the 26th International Joint Conference on Artificial Intelligence}}. \bibinfo{pages}{2464--2470}.
\newblock


\bibitem[Ouyang et~al\mbox{.}(2021)]%
        {ouyang2021learning}
\bibfield{author}{\bibinfo{person}{Wentao Ouyang}, \bibinfo{person}{Xiuwu Zhang}, \bibinfo{person}{Shukui Ren}, \bibinfo{person}{Li Li}, \bibinfo{person}{Kun Zhang}, \bibinfo{person}{Jinmei Luo}, \bibinfo{person}{Zhaojie Liu}, {and} \bibinfo{person}{Yanlong Du}.} \bibinfo{year}{2021}\natexlab{}.
\newblock \showarticletitle{Learning graph meta embeddings for cold-start ads in click-through rate prediction}. In \bibinfo{booktitle}{\emph{Proceedings of the 44th International ACM SIGIR Conference on Research and Development in Information Retrieval}}. \bibinfo{pages}{1157--1166}.
\newblock


\bibitem[Pan et~al\mbox{.}(2019)]%
        {pan2019warm}
\bibfield{author}{\bibinfo{person}{Feiyang Pan}, \bibinfo{person}{Shuokai Li}, \bibinfo{person}{Xiang Ao}, \bibinfo{person}{Pingzhong Tang}, {and} \bibinfo{person}{Qing He}.} \bibinfo{year}{2019}\natexlab{}.
\newblock \showarticletitle{Warm up cold-start advertisements: Improving ctr predictions via learning to learn id embeddings}. In \bibinfo{booktitle}{\emph{Proceedings of the 42nd International ACM SIGIR Conference on Research and Development in Information Retrieval}}. \bibinfo{pages}{695--704}.
\newblock


\bibitem[Pi et~al\mbox{.}(2020)]%
        {pi2020search}
\bibfield{author}{\bibinfo{person}{Qi Pi}, \bibinfo{person}{Guorui Zhou}, \bibinfo{person}{Yujing Zhang}, \bibinfo{person}{Zhe Wang}, \bibinfo{person}{Lejian Ren}, \bibinfo{person}{Ying Fan}, \bibinfo{person}{Xiaoqiang Zhu}, {and} \bibinfo{person}{Kun Gai}.} \bibinfo{year}{2020}\natexlab{}.
\newblock \showarticletitle{Search-based user interest modeling with lifelong sequential behavior data for click-through rate prediction}. In \bibinfo{booktitle}{\emph{Proceedings of the 29th ACM International Conference on Information \& Knowledge Management}}. \bibinfo{pages}{2685--2692}.
\newblock


\bibitem[Rendle(2010)]%
        {rendle2010factorization}
\bibfield{author}{\bibinfo{person}{Steffen Rendle}.} \bibinfo{year}{2010}\natexlab{}.
\newblock \showarticletitle{Factorization Machines}. In \bibinfo{booktitle}{\emph{Proceedings of the 2010 IEEE International Conference on Data Mining}}. \bibinfo{pages}{995--1000}.
\newblock


\bibitem[Rendle et~al\mbox{.}(2009)]%
        {rendle2009bpr}
\bibfield{author}{\bibinfo{person}{Steffen Rendle}, \bibinfo{person}{Christoph Freudenthaler}, \bibinfo{person}{Zeno Gantner}, {and} \bibinfo{person}{Lars Schmidt-Thieme}.} \bibinfo{year}{2009}\natexlab{}.
\newblock \showarticletitle{BPR: Bayesian personalized ranking from implicit feedback}. In \bibinfo{booktitle}{\emph{Proceedings of the Twenty-Fifth Conference on Uncertainty in Artificial Intelligence}}. \bibinfo{pages}{452--461}.
\newblock


\bibitem[Singh and Gordon(2008)]%
        {singh2008relational}
\bibfield{author}{\bibinfo{person}{Ajit~P Singh} {and} \bibinfo{person}{Geoffrey~J Gordon}.} \bibinfo{year}{2008}\natexlab{}.
\newblock \showarticletitle{Relational learning via collective matrix factorization}. In \bibinfo{booktitle}{\emph{Proceedings of the 14th ACM SIGKDD International Conference on Knowledge Discovery and Data Mining}}. \bibinfo{pages}{650--658}.
\newblock


\bibitem[Song et~al\mbox{.}(2019)]%
        {song2019autoint}
\bibfield{author}{\bibinfo{person}{Weiping Song}, \bibinfo{person}{Chence Shi}, \bibinfo{person}{Zhiping Xiao}, \bibinfo{person}{Zhijian Duan}, \bibinfo{person}{Yewen Xu}, \bibinfo{person}{Ming Zhang}, {and} \bibinfo{person}{Jian Tang}.} \bibinfo{year}{2019}\natexlab{}.
\newblock \showarticletitle{Autoint: Automatic feature interaction learning via self-attentive neural networks}. In \bibinfo{booktitle}{\emph{Proceedings of the 28th ACM International Conference on Information and Knowledge Management}}. \bibinfo{pages}{1161--1170}.
\newblock


\bibitem[Sun et~al\mbox{.}(2020)]%
        {sun2020multi}
\bibfield{author}{\bibinfo{person}{Rui Sun}, \bibinfo{person}{Xuezhi Cao}, \bibinfo{person}{Yan Zhao}, \bibinfo{person}{Junchen Wan}, \bibinfo{person}{Kun Zhou}, \bibinfo{person}{Fuzheng Zhang}, \bibinfo{person}{Zhongyuan Wang}, {and} \bibinfo{person}{Kai Zheng}.} \bibinfo{year}{2020}\natexlab{}.
\newblock \showarticletitle{Multi-modal knowledge graphs for recommender systems}. In \bibinfo{booktitle}{\emph{Proceedings of the 29th ACM International Conference on Information \& Knowledge Management}}. \bibinfo{pages}{1405--1414}.
\newblock


\bibitem[Sun et~al\mbox{.}(2021a)]%
        {sun2021fm2}
\bibfield{author}{\bibinfo{person}{Yang Sun}, \bibinfo{person}{Junwei Pan}, \bibinfo{person}{Alex Zhang}, {and} \bibinfo{person}{Aaron Flores}.} \bibinfo{year}{2021}\natexlab{a}.
\newblock \showarticletitle{FM2: Field-matrixed factorization machines for recommender systems}. In \bibinfo{booktitle}{\emph{Proceedings of the Web Conference 2021}}. \bibinfo{pages}{2828--2837}.
\newblock


\bibitem[Sun et~al\mbox{.}(2021b)]%
        {sun2021discerning}
\bibfield{author}{\bibinfo{person}{Ying Sun}, \bibinfo{person}{Hengshu Zhu}, \bibinfo{person}{Chuan Qin}, \bibinfo{person}{Fuzhen Zhuang}, \bibinfo{person}{Qing He}, {and} \bibinfo{person}{Hui Xiong}.} \bibinfo{year}{2021}\natexlab{b}.
\newblock \showarticletitle{Discerning decision-making process of deep neural networks with hierarchical voting transformation}.
\newblock \bibinfo{journal}{\emph{Advances in Neural Information Processing Systems}}  \bibinfo{volume}{34} (\bibinfo{year}{2021}), \bibinfo{pages}{17221--17234}.
\newblock


\bibitem[Sun et~al\mbox{.}(2024)]%
        {sun2024large}
\bibfield{author}{\bibinfo{person}{Ying Sun}, \bibinfo{person}{Hengshu Zhu}, \bibinfo{person}{Lu Wang}, \bibinfo{person}{Le Zhang}, {and} \bibinfo{person}{Hui Xiong}.} \bibinfo{year}{2024}\natexlab{}.
\newblock \showarticletitle{Large-scale online job search behaviors reveal labor market shifts amid COVID-19}.
\newblock \bibinfo{journal}{\emph{Nature Cities}} \bibinfo{volume}{1}, \bibinfo{number}{2} (\bibinfo{year}{2024}), \bibinfo{pages}{150--163}.
\newblock


\bibitem[Sun et~al\mbox{.}(2021c)]%
        {sun2021market}
\bibfield{author}{\bibinfo{person}{Ying Sun}, \bibinfo{person}{Fuzhen Zhuang}, \bibinfo{person}{Hengshu Zhu}, \bibinfo{person}{Qi Zhang}, \bibinfo{person}{Qing He}, {and} \bibinfo{person}{Hui Xiong}.} \bibinfo{year}{2021}\natexlab{c}.
\newblock \showarticletitle{Market-oriented job skill valuation with cooperative composition neural network}.
\newblock \bibinfo{journal}{\emph{Nature communications}} \bibinfo{volume}{12}, \bibinfo{number}{1} (\bibinfo{year}{2021}), \bibinfo{pages}{1992}.
\newblock


\bibitem[Tang and Wang(2018)]%
        {tang2018personalized}
\bibfield{author}{\bibinfo{person}{Jiaxi Tang} {and} \bibinfo{person}{Ke Wang}.} \bibinfo{year}{2018}\natexlab{}.
\newblock \showarticletitle{Personalized top-n sequential recommendation via convolutional sequence embedding}. In \bibinfo{booktitle}{\emph{Proceedings of the eleventh ACM International Conference on Web Search and Data Mining}}. \bibinfo{pages}{565--573}.
\newblock


\bibitem[Tian et~al\mbox{.}(2023)]%
        {tian2023eulernet}
\bibfield{author}{\bibinfo{person}{Zhen Tian}, \bibinfo{person}{Ting Bai}, \bibinfo{person}{Wayne~Xin Zhao}, \bibinfo{person}{Ji-Rong Wen}, {and} \bibinfo{person}{Zhao Cao}.} \bibinfo{year}{2023}\natexlab{}.
\newblock \showarticletitle{EulerNet: Adaptive Feature Interaction Learning via Euler's Formula for CTR Prediction}. In \bibinfo{booktitle}{\emph{Proceedings of the 46th International ACM SIGIR Conference on Research and Development in Information Retrieval}}.
\newblock


\bibitem[Volkovs et~al\mbox{.}(2017)]%
        {volkovs2017dropoutnet}
\bibfield{author}{\bibinfo{person}{Maksims Volkovs}, \bibinfo{person}{Guangwei Yu}, {and} \bibinfo{person}{Tomi Poutanen}.} \bibinfo{year}{2017}\natexlab{}.
\newblock \showarticletitle{DropoutNet: addressing cold start in recommender systems}. In \bibinfo{booktitle}{\emph{Proceedings of the 31st International Conference on Neural Information Processing Systems}}. \bibinfo{pages}{4964--4973}.
\newblock


\bibitem[Wang et~al\mbox{.}(2023)]%
        {wang2023fresh}
\bibfield{author}{\bibinfo{person}{Jianling Wang}, \bibinfo{person}{Haokai Lu}, \bibinfo{person}{Sai Zhang}, \bibinfo{person}{Bart Locanthi}, \bibinfo{person}{Haoting Wang}, \bibinfo{person}{Dylan Greaves}, \bibinfo{person}{Benjamin Lipshitz}, \bibinfo{person}{Sriraj Badam}, \bibinfo{person}{Ed~H Chi}, \bibinfo{person}{Cristos~J Goodrow}, {et~al\mbox{.}}} \bibinfo{year}{2023}\natexlab{}.
\newblock \showarticletitle{Fresh Content Needs More Attention: Multi-funnel Fresh Content Recommendation}. In \bibinfo{booktitle}{\emph{Proceedings of the 29th ACM SIGKDD Conference on Knowledge Discovery and Data Mining}}. \bibinfo{pages}{5082--5091}.
\newblock


\bibitem[Wang et~al\mbox{.}(2017)]%
        {wang2017deep}
\bibfield{author}{\bibinfo{person}{Ruoxi Wang}, \bibinfo{person}{Bin Fu}, \bibinfo{person}{Gang Fu}, {and} \bibinfo{person}{Mingliang Wang}.} \bibinfo{year}{2017}\natexlab{}.
\newblock \showarticletitle{Deep \& cross network for ad click predictions}.
\newblock In \bibinfo{booktitle}{\emph{Proceedings of the ADKDD'17}}. \bibinfo{pages}{1--7}.
\newblock


\bibitem[Wang et~al\mbox{.}(2021)]%
        {wang2021dcn}
\bibfield{author}{\bibinfo{person}{Ruoxi Wang}, \bibinfo{person}{Rakesh Shivanna}, \bibinfo{person}{Derek Cheng}, \bibinfo{person}{Sagar Jain}, \bibinfo{person}{Dong Lin}, \bibinfo{person}{Lichan Hong}, {and} \bibinfo{person}{Ed Chi}.} \bibinfo{year}{2021}\natexlab{}.
\newblock \showarticletitle{Dcn v2: Improved deep \& cross network and practical lessons for web-scale learning to rank systems}. In \bibinfo{booktitle}{\emph{Proceedings of the Web Conference 2021}}. \bibinfo{pages}{1785--1797}.
\newblock


\bibitem[Wei et~al\mbox{.}(2019)]%
        {wei2019mmgcn}
\bibfield{author}{\bibinfo{person}{Yinwei Wei}, \bibinfo{person}{Xiang Wang}, \bibinfo{person}{Liqiang Nie}, \bibinfo{person}{Xiangnan He}, \bibinfo{person}{Richang Hong}, {and} \bibinfo{person}{Tat-Seng Chua}.} \bibinfo{year}{2019}\natexlab{}.
\newblock \showarticletitle{MMGCN: Multi-modal graph convolution network for personalized recommendation of micro-video}. In \bibinfo{booktitle}{\emph{Proceedings of the 27th ACM International Conference on Multimedia}}. \bibinfo{pages}{1437--1445}.
\newblock


\bibitem[Wu et~al\mbox{.}(2022)]%
        {wu2022multi}
\bibfield{author}{\bibinfo{person}{Yiqing Wu}, \bibinfo{person}{Ruobing Xie}, \bibinfo{person}{Yongchun Zhu}, \bibinfo{person}{Xiang Ao}, \bibinfo{person}{Xin Chen}, \bibinfo{person}{Xu Zhang}, \bibinfo{person}{Fuzhen Zhuang}, \bibinfo{person}{Leyu Lin}, {and} \bibinfo{person}{Qing He}.} \bibinfo{year}{2022}\natexlab{}.
\newblock \showarticletitle{Multi-view multi-behavior contrastive learning in recommendation}. In \bibinfo{booktitle}{\emph{International conference on database systems for advanced applications}}. Springer, \bibinfo{pages}{166--182}.
\newblock


\bibitem[Xiao et~al\mbox{.}(2017)]%
        {xiao2017attentional}
\bibfield{author}{\bibinfo{person}{Jun Xiao}, \bibinfo{person}{Hao Ye}, \bibinfo{person}{Xiangnan He}, \bibinfo{person}{Hanwang Zhang}, \bibinfo{person}{Fei Wu}, {and} \bibinfo{person}{Tat-Seng Chua}.} \bibinfo{year}{2017}\natexlab{}.
\newblock \showarticletitle{Attentional factorization machines: learning the weight of feature interactions via attention networks}. In \bibinfo{booktitle}{\emph{Proceedings of the 26th International Joint Conference on Artificial Intelligence}}. \bibinfo{pages}{3119--3125}.
\newblock


\bibitem[Xin et~al\mbox{.}(2025)]%
        {xin2025llmcdsr}
\bibfield{author}{\bibinfo{person}{Haoran Xin}, \bibinfo{person}{Ying Sun}, \bibinfo{person}{Chao Wang}, {and} \bibinfo{person}{Hui Xiong}.} \bibinfo{year}{2025}\natexlab{}.
\newblock \showarticletitle{LLMCDSR: Enhancing Cross-Domain Sequential Recommendation with Large Language Models}.
\newblock \bibinfo{journal}{\emph{ACM Transactions on Information Systems}} (\bibinfo{year}{2025}).
\newblock


\bibitem[Yao et~al\mbox{.}(2021)]%
        {yao2021self}
\bibfield{author}{\bibinfo{person}{Tiansheng Yao}, \bibinfo{person}{Xinyang Yi}, \bibinfo{person}{Derek~Zhiyuan Cheng}, \bibinfo{person}{Felix Yu}, \bibinfo{person}{Ting Chen}, \bibinfo{person}{Aditya Menon}, \bibinfo{person}{Lichan Hong}, \bibinfo{person}{Ed~H Chi}, \bibinfo{person}{Steve Tjoa}, \bibinfo{person}{Jieqi Kang}, {et~al\mbox{.}}} \bibinfo{year}{2021}\natexlab{}.
\newblock \showarticletitle{Self-supervised learning for large-scale item recommendations}. In \bibinfo{booktitle}{\emph{Proceedings of the 30th ACM International Conference on Information \& Knowledge Management}}. \bibinfo{pages}{4321--4330}.
\newblock


\bibitem[Zhang et~al\mbox{.}(2016)]%
        {zhang2016collaborative}
\bibfield{author}{\bibinfo{person}{Fuzheng Zhang}, \bibinfo{person}{Nicholas~Jing Yuan}, \bibinfo{person}{Defu Lian}, \bibinfo{person}{Xing Xie}, {and} \bibinfo{person}{Wei-Ying Ma}.} \bibinfo{year}{2016}\natexlab{}.
\newblock \showarticletitle{Collaborative knowledge base embedding for recommender systems}. In \bibinfo{booktitle}{\emph{Proceedings of the 22nd ACM SIGKDD International Conference on Knowledge Discovery and Data Mining}}. \bibinfo{pages}{353--362}.
\newblock


\bibitem[Zhang et~al\mbox{.}(2023a)]%
        {zhang2023lightfr}
\bibfield{author}{\bibinfo{person}{Honglei Zhang}, \bibinfo{person}{Fangyuan Luo}, \bibinfo{person}{Jun Wu}, \bibinfo{person}{Xiangnan He}, {and} \bibinfo{person}{Yidong Li}.} \bibinfo{year}{2023}\natexlab{a}.
\newblock \showarticletitle{LightFR: Lightweight federated recommendation with privacy-preserving matrix factorization}.
\newblock \bibinfo{journal}{\emph{ACM Transactions on Information Systems}} \bibinfo{volume}{41}, \bibinfo{number}{4} (\bibinfo{year}{2023}), \bibinfo{pages}{1--28}.
\newblock


\bibitem[Zhang et~al\mbox{.}(2023b)]%
        {zhang2023interactive}
\bibfield{author}{\bibinfo{person}{He Zhang}, \bibinfo{person}{Ying Sun}, \bibinfo{person}{Weiyu Guo}, \bibinfo{person}{Yafei Liu}, \bibinfo{person}{Haonan Lu}, \bibinfo{person}{Xiaodong Lin}, {and} \bibinfo{person}{Hui Xiong}.} \bibinfo{year}{2023}\natexlab{b}.
\newblock \showarticletitle{Interactive interior design recommendation via coarse-to-fine multimodal reinforcement learning}. In \bibinfo{booktitle}{\emph{Proceedings of the 31st ACM International Conference on Multimedia}}. \bibinfo{pages}{6472--6480}.
\newblock


\bibitem[Zhang et~al\mbox{.}(2024)]%
        {zhang2024uncovering}
\bibfield{author}{\bibinfo{person}{Honglei Zhang}, \bibinfo{person}{Shuyi Wang}, \bibinfo{person}{Haoxuan Li}, \bibinfo{person}{Chunyuan Zheng}, \bibinfo{person}{Xu Chen}, \bibinfo{person}{Li Liu}, \bibinfo{person}{Shanshan Luo}, {and} \bibinfo{person}{Peng Wu}.} \bibinfo{year}{2024}\natexlab{}.
\newblock \showarticletitle{Uncovering the propensity identification problem in debiased recommendations}. In \bibinfo{booktitle}{\emph{2024 IEEE 40th International Conference on Data Engineering (ICDE)}}. IEEE, \bibinfo{pages}{653--666}.
\newblock


\bibitem[Zhang et~al\mbox{.}(2021a)]%
        {zhang2021multi}
\bibfield{author}{\bibinfo{person}{Kai Zhang}, \bibinfo{person}{Hao Qian}, \bibinfo{person}{Qing Cui}, \bibinfo{person}{Qi Liu}, \bibinfo{person}{Longfei Li}, \bibinfo{person}{Jun Zhou}, \bibinfo{person}{Jianhui Ma}, {and} \bibinfo{person}{Enhong Chen}.} \bibinfo{year}{2021}\natexlab{a}.
\newblock \showarticletitle{Multi-interactive attention network for fine-grained feature learning in ctr prediction}. In \bibinfo{booktitle}{\emph{Proceedings of the 14th ACM International Conference on Web Search and Data Mining}}. \bibinfo{pages}{984--992}.
\newblock


\bibitem[Zhang et~al\mbox{.}(2021b)]%
        {zhang2021talent}
\bibfield{author}{\bibinfo{person}{Qi Zhang}, \bibinfo{person}{Hengshu Zhu}, \bibinfo{person}{Ying Sun}, \bibinfo{person}{Hao Liu}, \bibinfo{person}{Fuzhen Zhuang}, {and} \bibinfo{person}{Hui Xiong}.} \bibinfo{year}{2021}\natexlab{b}.
\newblock \showarticletitle{Talent demand forecasting with attentive neural sequential model}. In \bibinfo{booktitle}{\emph{Proceedings of the 27th ACM SIGKDD Conference on Knowledge Discovery \& Data Mining}}. \bibinfo{pages}{3906--3916}.
\newblock


\bibitem[Zhou et~al\mbox{.}(2018)]%
        {zhou2018deep}
\bibfield{author}{\bibinfo{person}{Guorui Zhou}, \bibinfo{person}{Xiaoqiang Zhu}, \bibinfo{person}{Chenru Song}, \bibinfo{person}{Ying Fan}, \bibinfo{person}{Han Zhu}, \bibinfo{person}{Xiao Ma}, \bibinfo{person}{Yanghui Yan}, \bibinfo{person}{Junqi Jin}, \bibinfo{person}{Han Li}, {and} \bibinfo{person}{Kun Gai}.} \bibinfo{year}{2018}\natexlab{}.
\newblock \showarticletitle{Deep interest network for click-through rate prediction}. In \bibinfo{booktitle}{\emph{Proceedings of the 24th ACM SIGKDD International Conference on Knowledge Discovery \& Data Mining}}. \bibinfo{pages}{1059--1068}.
\newblock


\bibitem[Zhou et~al\mbox{.}(2020)]%
        {zhou2020s3}
\bibfield{author}{\bibinfo{person}{Kun Zhou}, \bibinfo{person}{Hui Wang}, \bibinfo{person}{Wayne~Xin Zhao}, \bibinfo{person}{Yutao Zhu}, \bibinfo{person}{Sirui Wang}, \bibinfo{person}{Fuzheng Zhang}, \bibinfo{person}{Zhongyuan Wang}, {and} \bibinfo{person}{Ji-Rong Wen}.} \bibinfo{year}{2020}\natexlab{}.
\newblock \showarticletitle{S3-rec: Self-supervised learning for sequential recommendation with mutual information maximization}. In \bibinfo{booktitle}{\emph{Proceedings of the 29th ACM International Conference on Information \& Knowledge Management}}. \bibinfo{pages}{1893--1902}.
\newblock


\bibitem[Zhu et~al\mbox{.}(2024)]%
        {zhu2024interest}
\bibfield{author}{\bibinfo{person}{Yongchun Zhu}, \bibinfo{person}{Jingwu Chen}, \bibinfo{person}{Ling Chen}, \bibinfo{person}{Yitan Li}, \bibinfo{person}{Feng Zhang}, {and} \bibinfo{person}{Zuotao Liu}.} \bibinfo{year}{2024}\natexlab{}.
\newblock \showarticletitle{Interest clock: Time perception in real-time streaming recommendation system}. In \bibinfo{booktitle}{\emph{Proceedings of the 47th International ACM SIGIR Conference on Research and Development in Information Retrieval}}. \bibinfo{pages}{2915--2919}.
\newblock


\bibitem[Zhu et~al\mbox{.}(2022)]%
        {zhu2022personalized}
\bibfield{author}{\bibinfo{person}{Yongchun Zhu}, \bibinfo{person}{Zhenwei Tang}, \bibinfo{person}{Yudan Liu}, \bibinfo{person}{Fuzhen Zhuang}, \bibinfo{person}{Ruobing Xie}, \bibinfo{person}{Xu Zhang}, \bibinfo{person}{Leyu Lin}, {and} \bibinfo{person}{Qing He}.} \bibinfo{year}{2022}\natexlab{}.
\newblock \showarticletitle{Personalized transfer of user preferences for cross-domain recommendation}. In \bibinfo{booktitle}{\emph{Proceedings of the Fifteenth ACM International Conference on Web Search and Data Mining}}. \bibinfo{pages}{1507--1515}.
\newblock


\bibitem[Zhu et~al\mbox{.}(2021)]%
        {zhu2021learning}
\bibfield{author}{\bibinfo{person}{Yongchun Zhu}, \bibinfo{person}{Ruobing Xie}, \bibinfo{person}{Fuzhen Zhuang}, \bibinfo{person}{Kaikai Ge}, \bibinfo{person}{Ying Sun}, \bibinfo{person}{Xu Zhang}, \bibinfo{person}{Leyu Lin}, {and} \bibinfo{person}{Juan Cao}.} \bibinfo{year}{2021}\natexlab{}.
\newblock \showarticletitle{Learning to warm up cold item embeddings for cold-start recommendation with meta scaling and shifting networks}. In \bibinfo{booktitle}{\emph{Proceedings of the 44th International ACM SIGIR Conference on Research and Development in Information Retrieval}}. \bibinfo{pages}{1167--1176}.
\newblock


\end{thebibliography}

\end{document}